\documentclass[a4paper]{ESASPCS13Style}
\usepackage{epsfig}

\begin{document}

\title{Dynamo action in late-type giants}

\author{S. B. F. Dorch\inst{1}}
  \institute{The Niels Bohr Institute for Astronomy, Physics and Geophysics,
     Juliane Maries Vej 30, DK-2100 Copenhagen {\O}, Denmark}

\maketitle

\begin{abstract}

Recent numerical MHD simulations suggest that magnetic activity
may occur in late-type giants. A entire red supergiant with
stellar parameters equal to Betelgeuse was modelled in 3d with the
high-order ``Pencil Code". Linear kinematic and non-linear
saturated dynamo action are found and the non-linear magnetic
field saturates at a super-equipartition value, while in the
linear regime two different modes of dynamo action are found.
Magnetic activity of late-type giants, if it exists, may influence
dust and wind formation and possibly lead to the heating of the
outer atmospheres of these stars.

\keywords{Stars: AGB, activity, Betelgeuse}

\end{abstract}

\section{Introduction}

Both recent theoretical and observational advances indicate that
late-type giant stars may possess magnetic fields. It has been
suggested that non-spherically symmetric planetary nebulae form
during late stages of AGB star evolution as a result of the
collimating effect of magnetic fields. From observations, maser
polarisation is known to exist in circumstellar envelopes of AGB
stars (e.g.\ Gray et al.\ \cite{Gray+ea99}, Vlemmings et al.\
\cite{Vlemmings+ea03}, and Sivagnanam \cite{Sivagnanam2004}) and
X-ray emission has been observed from some cool giant stars (e.g.\
H\"{u}nsch et al.\ \cite{Hunch+ea98} and Ayres et al.\
\cite{Ayres+ea03}). These observations are generally taken as
evidence for the existence of magnetic activity in late-type giant
stars (cf.\ Soker \& Kastner \cite{Soker+Kastner2003}).

The cool star Betelgeuse is a much observed late-type supergiant
that displays large-scale surface structures (e.g.\ Gray
\cite{Gray2000}). Freytag et al.\ (\cite{Freytag+ea02}) performed
detailed numerical 3-d simulations of the convective envelope of
the star under realistic physical assumptions, while trying to
determine if the star's known brightness fluctuations may be
understood as convective motions within the star's atmosphere: The
resulting models were successful in explaining the observations as
a consequence of giant-cell convection on the stellar surface.
Dorch \& Freytag (\cite{Dorch+Freytag2002}) performed a kinematic
dynamo analysis of the convective motions in the above model and
found that a weak seed magnetic field could indeed be
exponentially amplified by the giant-cell convection.

This is a report on recent full non-linear MHD simulations of
dynamo action in a late-type supergiant star with fundamental
stellar parameters set equal to that of Betelgeuse (see also Dorch
\cite{Dorch2004}).

\section{Model}

The full 3-d MHD equations was solved for a fully convective star
so that the entire star is contained within the computational box.
The computer code used is the ``Pencil Code'' by Brandenburg \&
Dobler\footnote{{\tt
http://www.nordita.dk/data/brandenb/pencil-code/}}, which has a
``convective star'' module that allows the solution of the
non-linear MHD equations by the numerical pencil scheme in a star
with a fixed radius R and mass M. The code solves the compressible
MHD equations and variables are computed in terms of R and M so
that the unit of the star's luminosity L becomes $\frac{\rm M}{\rm
R} ({\rm GM}/{\rm R})^{\frac{3}{2}}$. In the present case these
fundamental parameters are set to ${\rm R} = 640~ {\rm R}_{\odot}$
and ${\rm M} = 5~ {\rm M}_{\odot}$ yielding a luminosity of ${\rm
L} = 46000~ {\rm L}_{\odot}$, consistent with estimates of
Betelgeuse's size, mass and luminosity. The model uses a fixed
gravitational potential, an inner tiny heating core, and an outer
thin isothermally cooling spherical surface at $r = {\rm R}$, with
a cooling time scale set to $\tau_{\rm cool} = 1$ year
corresponding to the convective turn-over time in the model of
Freytag et al. (\cite{Freytag+ea02}).

Dynamo action by flows are sometimes studied in the limit of
increasingly large magnetic Reynolds numbers Re$_{\rm m} = \ell
{\rm U}/\eta$, where $\ell$ and U are characteristic length and
velocity scales, and $\eta$ is the magnetic diffusivity. Most
astrophysical systems are highly conducting resulting in a small
magnetic diffusivity $\eta$, and their sizes are huge yielding
enormously large values of Re$_{\rm m}$. Betelgeuse is not an
exception and most parts of the star are better conducting than
the solar photosphere that has $\eta \approx 10^4$ m$^2$/s: Taking
$\ell$ to be 10\% of the radial distance R from the centre, and U
to be an estimate of the RMS speed along the radius yields
Re$_{\rm m}=10^{10}$--$10^{12}$ in the interior of the star.

In the present case we cannot use that large values of Re$_{\rm
m}$, but rely on the results from generic dynamo experiments
showing that overall results converge already at Re$_{\rm m}$ of a
few hundred (e.g.\ Archontis, Dorch, \& Nordlund
\cite{Archontis+ea03a,Archontis+ea03b}). Here we use a value of
$\eta$ leading to a magnetic Reynolds number of Re$_{\rm m} \sim
300$.

\section{Discussion of results}

\begin{figure}[!htb]
\makebox[9.0cm]{ \epsfxsize=9.0cm \epsfysize=9.0cm
\epsfbox{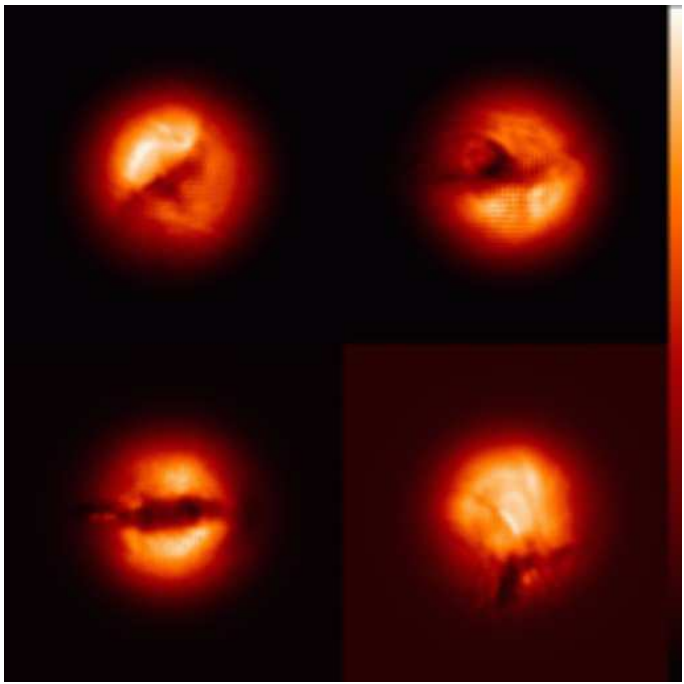} } \caption[]{Simulated surface intensity
snapshots at four different instants, time = 256, 347, 457 and 494
years (from upper left to lower right).} \label{fig1}
\end{figure}

Though it is not the topic of this contribution, it is appropriate
to discuss the properties of the convective flows in the model,
since these ultimately supply the kinetic energy forming the basic
energy reservoir for magnetic activity. It is not expected that
the flows match exactly what is found in more realistic radiation
simulations, but at least a qualitative agreement should be
inferred.

Large scale convection develop rapidly throughout the star and is
evident in several variables: However, since the model does not
incorporate realistic radiative transfer, only a simulated
intensity can be derived: Figure \ref{fig1} shows simulated
intensity snapshots at different instances. The contrast between
bright and dark patches on the surface is 20--50\%, and only 2--4
cells are seen at the stellar disk at any one time. The simulated
intensity is in qualitative agreement with the models of Freytag
et al.\ (\cite{Freytag+ea02}): The surface is not composed of
simply bright granules and dark intergranular lanes in the solar
sense---the large-scale convective patterns are typically larger
than 15--30\% of the radius, and are actually often on the order
of the radius in size. The corresponding radial velocities range
between 1--10 km/s in both up and down flowing regions.

\begin{figure}[!htb]
\makebox[8cm]{ \epsfxsize=8.0cm \epsfysize=6.0cm
\epsfbox{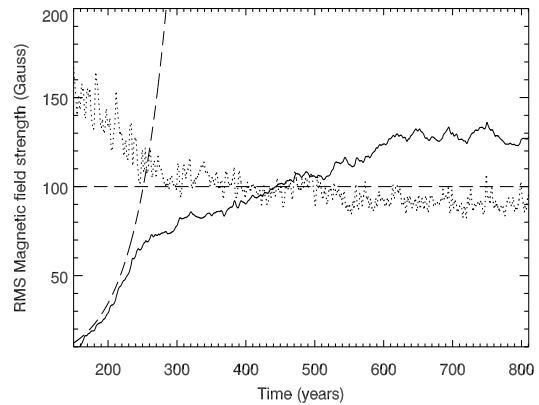} } \caption[]{Transition to the non-linear
regime: RMS magnetic field in the whole computational box as a
function of time in years. The upper curve is the equipartition
field strength corresponding to the average kinetic energy density
of the fluid motions (dotted curve) and the lower full curve is
the actual RMS field strength (full curve). The dashed thin curve
correspond to growth times of 25 years and a horizontal reference
line at 100 Gauss. } \label{fig2}
\end{figure}

Once the convection is taking place the magnetic field is
amplified and the system enters a linear regime of exponential
growth. There are two modes of amplification in the linear regime:
An initial mode with a growth rate of about 4 years, which in the
end gives way to a mode with a smaller growth rate corresponding
to a time-scale of about 25 years.

\begin{figure}[!htb]
\makebox[8cm]{ \epsfxsize=8.0cm \epsfysize=6.0cm
\epsfbox{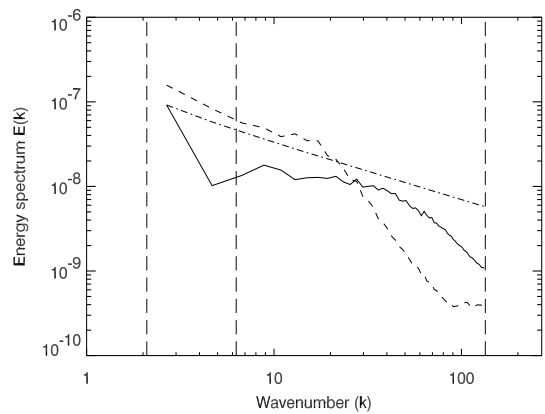} } \caption[]{Energy spectra of the
magnetic energy density (solid curve) and kinetic energy density
(dashed curve) at time = 512 years, and a line corresponding to a
power-law with an exponent of -2/3 (dashed-dotted line). The
vertical lines, from left to right, denote the wavenumbers
corresponding to the computational box size, the stellar radius,
and the numerical resolution. } \label{fig3}
\end{figure}

No exponential growth can go on forever and eventually the
magnetic energy amplification must stop: The question is whether
the magnetic field retains a roughly constant saturation value, or
if it starts to dissipate. In case of saturation the typical field
strength is likely to be on the order of the equipartition value
corresponding to equal magnetic and kinetic energy densities.
Figure \ref{fig2} shows the magnetic field strength B$_{\rm RMS}$
within the entire star as a function of time for $\sim 800$ years:
The second linear mode as well as the mode in the non-linear
regime are visible. The magnetic field saturates at a value
slightly above the equipartition field strength B$_{\rm eq} =
\sqrt{\mu_0 \left< \rho {\rm u}^2 \right>}\sim$ 90--100 Gauss,
corresponding to a value of about 120--130 Gauss. In terms of
total energy this corresponds to magnetic energy E$_{\rm mag}$
being above equipartition with the kinetic energy E$_{\rm kin}$ by
roughly a factor of two. Hence the field is not extremely strong,
but is neither particularly weak. From the observational view
point the strength of the field at the surface is interesting. In
the non-linear regime the field strength at the sphere at $r =
{\rm R}$ can be up to $\sim 500$ Gauss, while in the interior of
the star the field strength rises and can be as high as a few kG:
The strongest magnetic structures completely quenches the velocity
field in these regions, i.e.\ the local field can be far above
equipartition. The downward increase of the field strength is
similar to the flux pumping effect found in the solar context
(cf.\ Dorch \& Nordlund \cite{Dorch+Nordlund2001}).

\begin{figure}[!htb]
\makebox[8.75cm]{ \epsfxsize=8.75cm \epsfysize=8cm
\epsfbox{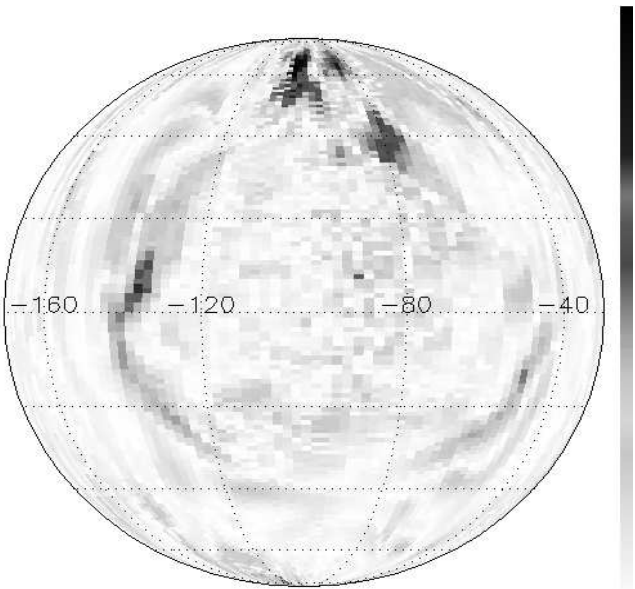} } \caption[]{An illustration of the
unsigned magnetic field strength at the spherical surface $r =$ R
of the model star as an orthographic map. The darkest patches
correspond to a maximum field strength of 500 Gauss. From a
snapshot at time = 695 years. The views are centered on a
longitude of $-100^{\rm o}$. The grid indicated has a longitudinal
spacing of $40^{\rm o}$ and a latitudinal spacing of $20^{\rm
o}$.} \label{fig4}
\end{figure}

The geometry of the magnetic field in the saturating non-linear
stage of the dynamo may since this could be relevant for the
influence of the field on the formation of asymmetric dust and
wind: The field is concentrated into elongated relatively thin
structures, but formes no intergranular network.

Energy spectra reveal that the magnetic structures are well
resolved, and that the power at the largest wavenumbers $k\sim100$
is a few orders of magnitude smaller than that at the largest
scales (see Figure \ref{fig3}). The power is maximum on the
largest scales corresponding to wavenumbers of a few, while there
is a dip at $k\sim7$ corresponding to the scale of the radius,
where the power is minimum. The power on scales $k\approx$ 10--20
is roughly proportional to $k^{-2/3}$ corresponding to Kolmogorov
scaling, and at small-scales $k\ga50$ the power steeply drops:
Magnetic structures in the non-linear regime are then large by
solar standards, but smaller than the convection cells.

\begin{figure}[!htb]
\makebox[8.75cm]{ \epsfxsize=8.75cm \epsfysize=8cm
\epsfbox{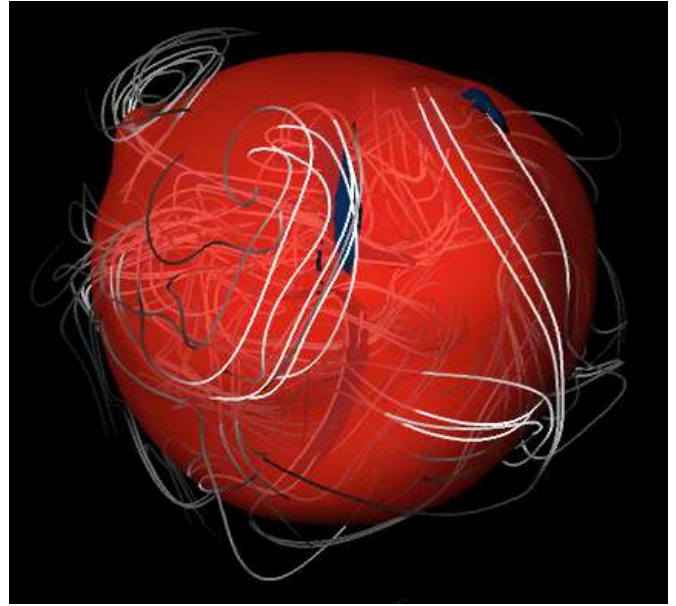} } \caption[]{A 3-d volume rendering of
magnetic field lines (white) and flux ropes (dark structures).
Also show is a isosurface (transparent) at the surface temperature
value.} \label{fig5}
\end{figure}

Figure \ref{fig4} is a map of the spherical surface at $r =$ R:
There are both dark patches of strong magnetic field and large
areas with a vanishing field. On this map the areal magnetic
filling factor is $\sim$55\% for $|{\rm B}|>50$ Gauss, while it is
only $\sim$0.6\% for $|{\rm B}|>500$ Gauss. This map, however,
does not represent a physical surface of the star. Due to the fact
that the actual upper boundary consists of a few large cells the
surface cannot be captured by a sphere with radius R. This fact is
demonstrated by a volume rendering (Fig.\ \ref{fig5}) of the
isosurface at the cooling temperature: In Figure \ref{fig5} there
are several slopes in the temperature isosurface.

\section{Summary and conclusion}

Three different modes of dynamo action are recognised: A fast
growing linear mode with a growth time of $\sim 4$ years, a
relatively slowly growing mode with an exponential growth of $\sim
25$ years, and finally a saturated non-linear mode operating a
factor of two above equipartition. More modes may exist, but they
must have very low growth rates or very small amplitudes since
they have not appeared in the simulations.

On the one hand, it is not possible to state conclusively if
Betelgeuse actually has a magnetic field, on the basis of the
present model, since such a field is unobserved. On the other
hand, one can state that it seems possible that late-type giant
stars such as Betelgeuse may have presently undetected magnetic
fields. These fields are possibly close to equipartition yielding
surface strengths up to 500 Gauss. The latter field strengths may
be difficult to detect directly, due to the relatively small
filling factors of the strong fields, but even the moderately
strong fields may influence their immediate surroundings.

Finally it is note worthy that Lobel et al.\ (\cite{Lobel+ea04})
recently published spatially resolved spectra of the upper
chromosphere and dust envelope of Betelgeuse. These observations
reveal that the chromosphere extends far beyond the circumstellar
envelope. The presence of a hot or warm chromosphere may lead one
to speculate on the possible connection to coronal heating in
solar-like stars, which is magnetic in origin and caused by flux
braiding motions in the photosphere (Gudiksen \& Nordlund
\cite{Gudiksen+Nordlund2002}): It remains to be studied whether a
similar process could be taking place in late-type giants.

\begin{acknowledgements}
This work was supported by the Danish Natural Science Research
Council. Access to computational resources granted by the Danish
Center for Scientific Computing in particular the Horseshoe
cluster at Odense University.
\end{acknowledgements}

\end{document}